\documentclass{svmult}

\usepackage{mathptmx} 
\usepackage{helvet}
\usepackage{courier}
\usepackage{graphicx}
\usepackage[bottom]{footmisc}
\usepackage{amsmath,amssymb,amsfonts}
\usepackage{float}
\usepackage{cite}

\def\Ei{{\rm Ei}}
\def\rd{{\rm d}}
\def\xcrit{x_{\rm crit}}

\begin{document}

\title*{The effect of a pressure-dependent viscosity on the viscous scraper problem}
\author{Fiaz Ur Rehman and Stephen K.\ Wilson}
\institute{
Fiaz Ur Rehman \at School of Mathematics, Monash University, Victoria 3800, Australia \email{fiaz.urrehman@monash.edu}
\and 
Stephen K.\ Wilson \at Department of Mathematical Sciences, University of Bath, Bath BA2 7AY, United Kingdom \email{sw3197@bath.ac.uk}
}
\maketitle
\abstract{
The effect of a pressure-dependent viscosity on the behaviour of the viscous scraper problem is investigated.
In particular, it is found that the effect is qualitatively different for the classical scraper (i.e., a drag in) problem and the reverse scraper (i.e., a drag out) problem.
}

% Section 1
\section{Introduction}
\label{sec:sec1}

In extreme physical conditions the viscosity of a Newtonian fluid can increase significantly with pressure, a phenomenon known as piezoviscosity (see, for example, \cite{Herbst,Bair,Barus}).
In this work, we investigate the effect of a pressure-dependent viscosity on the behaviour of a paradigm lubrication flow in which high pressures can occur, namely the viscous scraper problem (see, for example, \cite{Batchelor}).

% Section 1
\section{The Viscous Scraper Problem}
\label{sec:sec2}

% Figure 1
\begin{figure}[t]
\centering
\includegraphics[scale=0.9]{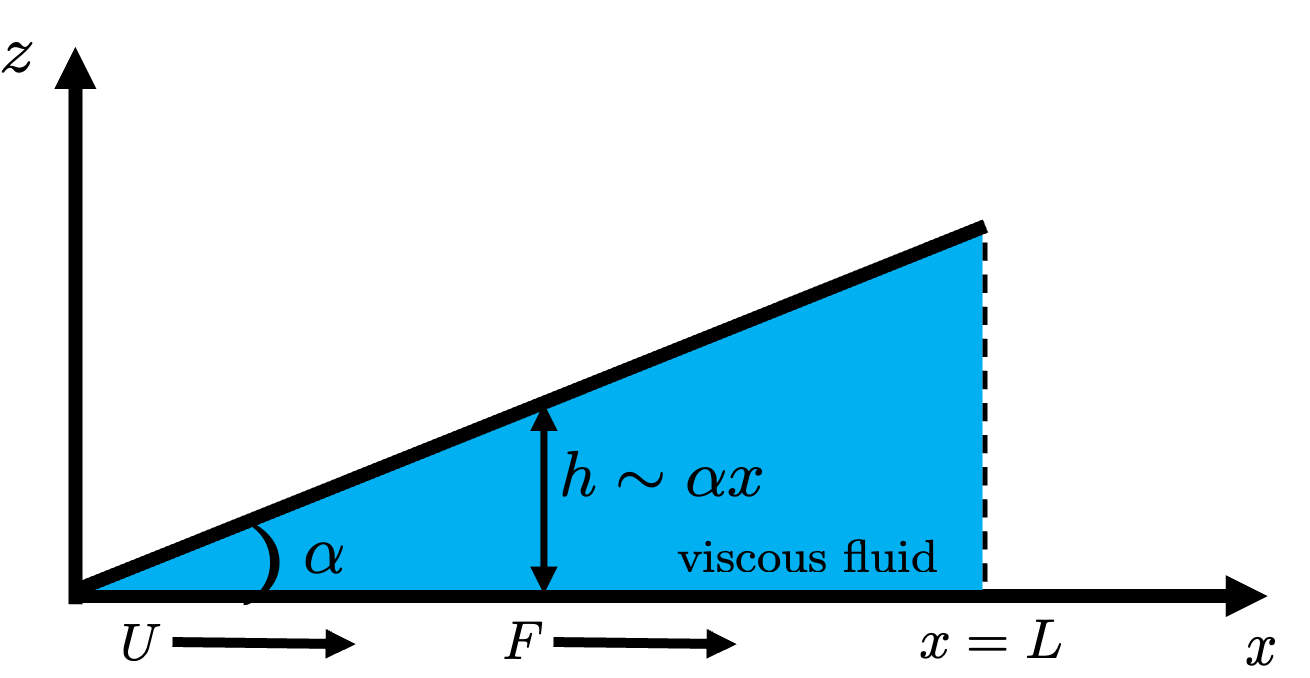} 
\caption{Schematic of the viscous scraper problem. The case $U<0$ corresponds to the classical scraper (i.e., a drag in) problem and the case $U>0$ corresponds to the reverse scraper (i.e., a drag out) problem.
}
\label{fig:1}
\end{figure}

Consider the two-dimensional flow of a viscous fluid between two plates referred to a Cartesian coordinate system $(x,z)$.
The lower plate is at $z=0$ and the upper plate is at $z=h(x)=x\tan{\alpha}$ for $0 \le x \le L$ and so is of length $L$ and makes an angle $\alpha\,(>0)$ with the lower plate.
The upper plate is stationary and the lower plate moves in the positive $x$ direction with constant speed $U$, with $U<0$ (i.e., moving to the left) corresponding to the classical viscous scraper (i.e., a drag in) problem and $U>0$ (i.e., moving to the right) corresponding to the reverse viscous scraper (i.e., a drag out) problem.
A schematic of the problem is shown in Fig.\ 1.

We model the fluid within the gap between the plates using the incompressible continuity and the Navier--Stokes equations,
\begin{equation}\label{eq:eq1}
    \nabla\cdot\textbf{u}=0
\end{equation}
and
\begin{equation}\label{eq:eq2}
    \rho\left(\frac{\partial \textbf{u}}{\partial t}+\textbf{u}\cdot\nabla\textbf{u}\right)=-\nabla p+\nabla\cdot\left(\mu\left(\nabla\textbf{u}+\nabla\textbf{u}^T\right)\right)-\rho\textbf{g},
\end{equation}
where $\textbf{u}(x,z,t)=(u(x,z,t),w(x,z,t))$ is the velocity, $p(x,z,t)$ is the pressure, $\mu=\mu(p)$ is the pressure-dependent viscosity, $\rho$ is the constant density, $\textbf{g}$ is constant gravitational acceleration acting vertically downwards (i.e., in the negative $z$ direction), and $t$ denotes time.

For simplicity, we consider steady lubrication (i.e., thin-film) flow in the absence of gravity, in which case $\alpha\ll 1$ and hence $h \sim \alpha x$ and (\ref{eq:eq2}) reduces to the lubrication equations,
\begin{equation}\label{eq:lub1}
    \frac{\partial}{\partial z}\left(\mu\frac{\partial u}{\partial z}\right)=\frac{\partial p}{\partial x}
\end{equation}
and
\begin{equation}\label{eq:lub2}
    \frac{\partial p}{\partial z}=0,
\end{equation}
subject to no-slip and no-penetration boundary conditions at the plates,
\begin{equation}\label{eq:bc1}
    u=U, \quad w=0 \quad \text{at} \quad z=0,
\end{equation}
\begin{equation}\label{eq:bc2}
    u=0, \quad w=0 \quad \text{at} \quad z=h. 
\end{equation} 
From $(\ref{eq:lub2})$ we deduce that $p$, and hence $\mu$, is independent of $z$, i.e., $p$ and $\mu$ are functions of only $x$.
Hence for any viscosity $\mu=\mu(p)$ we can integrate $(\ref{eq:lub1})$ twice with respect to $z$ to find the horizontal velocity of the fluid subject to the boundary conditions (\ref{eq:bc1}) and (\ref{eq:bc2}) to be
\begin{equation}\label{eq:u1}
    u=-\frac{1}{2\mu}\frac{\rd p}{\rd x}z(h-z)+U\left(1-\frac{z}{h}\right).
\end{equation}
The volume flux (per unit width) of the fluid within the gap between the plates is given by
\begin{equation}\label{eq:Q1}
    Q=\int_0^hu\,\rd z.
\end{equation}
Substituting the expression for $u$ given by $(\ref{eq:u1})$ into $(\ref{eq:Q1})$ and performing the integration yields
\begin{equation}\label{eq:Q2}
    Q=-\frac{h^3}{12\mu}\frac{\rd p}{\rd x}+\frac{Uh}{2}.
\end{equation}
For a steady flow, $Q$ is constant, and, since there is no flow through $x=0$, 
\begin{equation}\label{eq:Q3}
    Q=0.
\end{equation}
Combining (\ref{eq:Q2}) and (\ref{eq:Q3}) we obtain
\begin{equation}\label{eq:dpdx}
    \frac{\rd p}{\rd x}=\frac{6\mu U}{h^2},
\end{equation}
and substituting (\ref{eq:dpdx}) into (\ref{eq:u1}) yields
\begin{equation}\label{eq:u2}
    u=\frac{U(h-z)(h-3z)}{h^2}
\end{equation}
and hence
\begin{equation}
    \mu\frac{\partial u}{\partial z}=-\frac{2\mu U(2h-3z)}{h^2}.
\end{equation}
Since $h \sim \alpha x$, equation (\ref{eq:dpdx}) can be written as
\begin{equation}\label{eq:reynolds}
    \frac{\rd p}{\rd x}=\frac{6\mu U}{\alpha^2 x^2},
\end{equation}
which is a first-order ordinary differential equation for the pressure within the gap between the plates whose solution depends on the manner in which the viscosity depends on the pressure.
In order to determine $p$ we impose the condition that it takes the prescribed constant ambient value $p_L$ at the right-hand end of the gap, i.e.,
\begin{equation}\label{eq:bc3}
    p=p_L \quad \hbox{at} \quad x=L.
\end{equation}

The local shear stress in the positive $x$ direction (per unit width) exerted by the fluid on the lower plate is given by
\begin{equation}\label{eq:tau}
    \tau=-\mu\left.\frac{\partial u}{\partial z}\right|_{z=0}=\frac{4\mu U}{\alpha x},
\end{equation}
and hence the total force in the positive $x$ direction (per unit width) required to move the lower plate is given by
\begin{equation}\label{eq:F}
    F=\lim_{x \to 0^+}\int_x^L\tau\,\rd x=\lim_{x \to 0^+}\int_x^L\frac{4\mu U}{\alpha x}\,\rd x,
\end{equation}
both of which also depend on the manner in which the viscosity depends on the pressure. 

In what follows we consider three simple models for the dependence of $\mu$ on $p$, all of which satisfy
\begin{equation}
    \mu=\mu_0 \quad \hbox{and} \quad \frac{\rd \mu}{\rd p}=\beta \quad \hbox{at} \quad  p=0,
\end{equation}
where $\mu_0$ and $\beta$ are non-negative constants.

% Section 2.1
\subsection{Constant Viscosity $\mu=\mu_0$}
\label{sec:sec2_1}

For reference, we first consider the well-known case of a fluid with constant viscosity given by
\begin{equation}\label{eq:mu1}
    \mu=\mu_0,
\end{equation}
corresponding to $\beta=0$.
Solving (\ref{eq:reynolds}) with (\ref{eq:mu1}) subject to (\ref{eq:bc3}) yields the solution for the pressure, namely
\begin{equation}\label{eq:p1}
    p=p_L-\frac{6\mu_0U}{\alpha^2}\left(\frac{1}{x}-\frac{1}{L}\right),
\end{equation}
and hence the local shear stress is given by
\begin{equation}\label{eq:tau1}
    \tau=\frac{4\mu_0U}{\alpha x}
\end{equation}
and the total force in given by
\begin{equation}\label{eq:F1}
    F=\lim_{x \to 0^+}\frac{4\mu_0U}{\alpha}\log\left(\frac{L}{x}\right).
\end{equation}
In particular, the pressure is singular like
\begin{equation}\label{eq:psing1}
    p\sim-\frac{6\mu_0U}{\alpha^2 x} \to 
    \left\{
    \begin{matrix}
    +\infty \quad & \hbox{for} \quad U < 0 \\
    -\infty \quad & \hbox{for} \quad U > 0 \\
    \end{matrix}
    \right\}
\end{equation}
and the local shear stress is singular like
\begin{equation}\label{eq:tausing1}
    \tau\sim\frac{4\mu_0U}{\alpha x} \to 
    \left\{
    \begin{matrix}
    -\infty \quad & \hbox{for} \quad U < 0 \\
    +\infty \quad & \hbox{for} \quad U > 0 \\
    \end{matrix}
    \right\}
\end{equation}
as $x \to 0^+$,
and hence the total force is singular like
\begin{equation}\label{eq:Fsing1}
    F\sim\frac{4\mu_0U}{\alpha}(-\log x) \to 
    \left\{
    \begin{matrix}
    -\infty \quad & \hbox{for} \quad U < 0 \\
    +\infty \quad & \hbox{for} \quad U > 0 \\
    \end{matrix}
    \right\}
\end{equation}
as $x \to 0^+$,
and so the non-integrable singularity in $\tau$ as $x \to 0^+$ means that,
at least in theory,
an infinite force is required to move the lower plate.

% Section 2.2
\subsection{Linear Viscosity $\mu=\beta p$}
\label{sec:sec2_2}

For a pressure-dependent viscosity of the form
\begin{equation}
    \mu=\beta p,
\end{equation}
corresponding to $\mu_0=0$,
the pressure satisfies the separable differential equation
\begin{equation}
    \frac{1}{p}\frac{\rd p}{\rd x}=\frac{6\beta U}{\alpha^2 x^2}
\end{equation}
subject to (\ref{eq:bc3}), and so is given by 
\begin{equation}\label{eq:p2}
    p=p_L\exp\left(-\frac{6\beta U}{\alpha^2}\left(\frac{1}{x}-\frac{1}{L}\right)\right),
\end{equation}
and hence the viscosity is given by
\begin{equation}\label{eq:mu2}
    \mu=\beta p_L\exp\left(-\frac{6\beta U}{\alpha^2}\left(\frac{1}{x}-\frac{1}{L}\right)\right),
\end{equation}
the local shear stress is given by
\begin{equation}\label{eq:tau2}
    \tau=\frac{4\beta U}{\alpha x}p_L\exp\left(-\frac{6\beta U}{\alpha^2}\left(\frac{1}{x}-\frac{1}{L}\right)\right),
\end{equation}
and the total force in given by
\begin{equation}\label{eq:F2}
   F=\lim_{x \to 0^+}\frac{4\beta U}{\alpha}p_L\exp\left({\frac{6\beta U}{\alpha^2 L}}\right)\left[\Ei\left(-\frac{6\beta U}{\alpha^2 x}\right)-\Ei\left(-\frac{6\beta U}{\alpha^2 L}\right)\right],
\end{equation}
where $\Ei(\cdot)$ is the exponential integral function defined by
\begin{equation}
\Ei(x)=
\int_{-\infty}^{x} \frac{\exp(k)}{k}\,\rd k,
\end{equation}
which satisfies
\begin{equation}
    \Ei(x)\sim\frac{\exp(x)}{x} \to +\infty
    \quad \hbox{as} \quad x \to +\infty,
\end{equation}
\begin{equation}
    \Ei(x)=\log\vert x \vert + \gamma + x + O(x^2)
    \quad \hbox{as} \quad x \to 0
\end{equation}
and
\begin{equation}
    \Ei(x) \sim \frac{\exp(x)}{x} \to 0^-
    \quad \hbox{as} \quad x \to -\infty.
\end{equation}
In particular, the pressure behaves like
\begin{equation}\label{eq:psing2}
    p\sim p_L\exp\left(-\frac{6\beta U}{\alpha^2 x}\right) \to
    \left\{
    \begin{matrix}
    +\infty \quad & \hbox{for} \quad U < 0 \\
    0^+     \quad & \hbox{for} \quad U > 0 \\
    \end{matrix}
    \right\},
\end{equation}
the viscosity behaves like
\begin{equation}\label{eq:musing2}
    \mu\sim\beta p_L\exp\left(-\frac{6\beta U}{\alpha^2 x}\right) \to
    \left\{
    \begin{matrix}
    +\infty \quad & \hbox{for} \quad U < 0 \\
    0^+     \quad & \hbox{for} \quad U > 0 \\
    \end{matrix}
    \right\},
\end{equation}
and the local shear stress behaves like
\begin{equation}\label{eq:tausing2}
    \tau\sim\frac{4\beta U}{\alpha x}p_L\exp\left(-\frac{6\beta U}{\alpha^2 x}\right) \to
    \left\{
    \begin{matrix}
    -\infty \quad & \hbox{for} \quad U < 0 \\
    0^+     \quad & \hbox{for} \quad U > 0 \\
    \end{matrix}
    \right\}
\end{equation}
as $x \to 0^+$, 
and hence the total force behaves like
\begin{equation}\label{eq:Fsing2_1}
    F\sim-\frac{2\alpha x}{3}p_L\exp\left({\frac{6\beta U}{\alpha^2 L}}\right)\exp\left(-\frac{6\beta U}{\alpha^2 x}\right)
    \to -\infty 
    \quad \hbox{for} \quad U < 0
\end{equation}
as $x \to 0^+$ and is given by
\begin{equation}\label{eq:Fsing2_2}
    F=-\frac{4\beta U}{\alpha}p_L\exp\left({\frac{6\beta U}{\alpha^2 L}}\right)\Ei\left(-\frac{6\beta U}{\alpha^2 L}\right)
    \quad \hbox{for} \quad U > 0.
\end{equation}

% Section 2.3
\subsection{Linear Viscosity $\mu=\mu_0(1+\beta p/\mu_0)$}
\label{sec:sec2_3}

For a pressure-dependent viscosity of the form
\begin{equation}
    \mu=\mu_0\left(1+\frac{\beta p}{\mu_0}\right),
\end{equation}
the pressure satisfies the separable differential equation
\begin{equation}
   \frac{1}{1+\frac{\beta p}{\mu_0}}\frac{\rd p}{\rd x}=\frac{6\mu_0U}{\alpha^2 x^2},
\end{equation}
subject to (\ref{eq:bc3}), and so is given by
\begin{equation}\label{eq:p3}
    p=\frac{\mu_0}{\beta}\left[\left(1+\frac{\beta p_L}{\mu_0}\right)\exp\left(-\frac{6\beta U}{\alpha^2}\left(\frac{1}{x}-\frac{1}{L}\right)\right)-1\right],
\end{equation}
and hence the viscosity is given by
\begin{equation}\label{eq:mu3}
    \mu=\mu_0\left(1+\frac{\beta p_L}{\mu_0}\right)\exp\left(-\frac{6\beta U}{\alpha^2}\left(\frac{1}{x}-\frac{1}{L}\right)\right),
\end{equation}
the local shear stress is given by
\begin{equation}\label{eq:tau3}
    \tau=\frac{4\mu_0 U}{\alpha x}\left(1+\frac{\beta p_L}{\mu_0}\right)\exp\left(-\frac{6\beta U}{\alpha^2}\left(\frac{1}{x}-\frac{1}{L}\right)\right),
\end{equation}
and hence the total force is given by
\begin{equation}\label{eq:F3}
    F=\lim_{x \to 0^+}\frac{4\mu_0U}{\alpha}\left(1+\frac{\beta p_L}{\mu_0}\right)\exp\left(\frac{6\beta U}{\alpha^2 L}\right)\left[\Ei\left(-\frac{6\beta U}{\alpha^2 x}\right)-\Ei\left(-\frac{6\beta U}{\alpha^2 L}\right)\right].
\end{equation}
In particular, the pressure behaves like
\begin{equation}\label{eq:psing3}
    p\sim\frac{\mu_0}{\beta}\left[\left(1+\frac{\beta p_L}{\mu_0}\right)\exp\left(-\frac{6\beta U}{\alpha^2 x}\right)-1\right] \to
    \left\{
    \begin{matrix}
    +\infty \quad & \hbox{for} \quad U < 0 \\
    -\frac{\mu_0}{\beta}^+ \quad & \hbox{for} \quad U > 0 \\
    \end{matrix}
    \right\},
\end{equation}
the viscosity behaves like
\begin{equation}\label{eq:musing3}
    \mu\sim\mu_0\left(1+\frac{\beta p_L}{\mu_0}\right)\exp\left(-\frac{6\beta U}{\alpha^2 x}\right) \to
    \left\{
    \begin{matrix}
    +\infty \quad & \hbox{for} \quad U < 0 \\
    0^+     \quad & \hbox{for} \quad U > 0 \\
    \end{matrix}
    \right\},
\end{equation}
and the local shear stress behaves like
\begin{equation}\label{eq:tausing3}
    \tau\sim\frac{4\mu_0 U}{\alpha x}\left(1+\frac{\beta p_L}{\mu_0}\right)\exp\left(-\frac{6\beta U}{\alpha^2 x}\right) \to
    \left\{
    \begin{matrix}
    -\infty \quad & \hbox{for} \quad U < 0 \\
    0^+     \quad & \hbox{for} \quad U > 0 \\
    \end{matrix}
    \right\}
\end{equation}
as $x \to 0^+$,
and hence the total force behaves like
\begin{equation}\label{eq:Fsing3_1}
    F\sim-\frac{2\mu_0\alpha x}{3\beta}\left(1+\frac{\beta p_L}{\mu_0}\right)\exp\left({\frac{6\beta U}{\alpha^2 L}}\right)\exp\left(-\frac{6\beta U}{\alpha^2 x}\right)
    \to -\infty 
    \quad \hbox{for} \quad U < 0
\end{equation}
as $x \to 0^+$ and is given by
\begin{equation}\label{eq:Fsing3_2}
    F=-\frac{4\mu_0 U}{\alpha}\left(1+\frac{\beta p_L}{\mu_0}\right)\exp\left({\frac{6\beta U}{\alpha^2 L}}\right)\Ei\left(-\frac{6\beta U}{\alpha^2 L}\right)
    \quad \hbox{for} \quad U > 0.
\end{equation}

% Section 2.4
\subsection{Exponential Viscosity $\mu=\mu_0\exp({\beta p}/{\mu_0})$}
\label{sec:sec2_4}

For a pressure-dependent viscosity of the form
\begin{equation}
    \mu=\mu_0\exp\left(\frac{\beta p}{\mu_0}\right),
\end{equation}
the pressure satisfies the separable differential equation
\begin{equation}
    \exp\left(-\frac{\beta p}{\mu_0}\right)\frac{\rd p}{\rd x}=\frac{6\mu_0 U}{\alpha^2 x^2}
\end{equation}
subject to (\ref{eq:bc3}), and so is given by
\begin{equation}\label{eq:p4}
    p=-\frac{\mu_0}{\beta}\log\left[\exp\left(-\frac{\beta p_L}{\mu_0}\right)+\frac{6\beta U}{\alpha^2}\left(\frac{1}{x}-\frac{1}{L}\right)\right],
\end{equation}
and hence the viscosity is given by
\begin{equation}\label{eq:mu4}
    \mu=\mu_0\left[\exp\left(-\frac{\beta p_L}{\mu_0}\right)+\frac{6\beta U}{\alpha^2}\left(\frac{1}{x}-\frac{1}{L}\right)\right]^{-1},
\end{equation}
the local shear stress is given by
\begin{equation}\label{eq:tau4}
    \tau=\frac{4\mu_0 U}{\alpha x}\left[\exp\left(-\frac{\beta p_L}{\mu_0}\right)+\frac{6\beta U}{\alpha^2}\left(\frac{1}{x}-\frac{1}{L}\right)\right]^{-1},
\end{equation}
and hence the total force is given by
\begin{eqnarray}\label{eq:F4}
    F=\lim_{x \to 0^+}\frac{4\mu_0U}{\alpha}\left\{\exp\left(-\frac{\beta p_L}{\mu_0}\right)-\frac{6\beta U}{\alpha^2 L}\right\}^{-1} \nonumber \\
    \times\log\left[\frac{L\exp\left(-\frac{\beta p_L}{\mu_0}\right)}{\left\{\exp\left(-\frac{\beta p_L}{\mu_0}\right)-\frac{6\beta U}{\alpha^2 L}\right\}x+\frac{6\beta U}{\alpha^2}}\right].
\end{eqnarray}
Note that in the case $U<0$ the expressions $p$, $\mu$ and $\tau$ given by (\ref{eq:p4}), (\ref{eq:mu4}) and (\ref{eq:tau4}), respectively, exist only in the interval $\xcrit < x \le L$, where $\xcrit$ $(0 < \xcrit < L)$ is defined by
\begin{equation}
    \xcrit=-\frac{6\beta U}{\alpha^2}\left\{\exp\left(-\frac{\beta p_L}{\mu_0}\right)-\frac{6\beta U}{\alpha^2 L}\right\}^{-1},
\end{equation}
and so in this case the expression for the total force on the lower plate given by (\ref{eq:F4}) is replaced by the corresponding expression for the total force on the portion of the lower plate lying in the interval $\xcrit < x \le L$ given by replacing the limit $x \to 0^+$ with the limit $x \to \xcrit^+$.

In the case $U<0$, the pressure is singular like
\begin{equation}\label{eq:psing4_1}
   p\sim-\frac{\mu_0}{\beta}\log\left[-\frac{6\beta U}{\alpha^2}\left(\frac{1}{\xcrit}-\frac{1}{x}\right)\right] \to +\infty,
\end{equation}
the viscosity is singular like
\begin{equation}\label{eq:musing4_1}
    \mu\sim-\frac{\mu_0\alpha^2}{6\beta U}\left(\frac{1}{\xcrit}-\frac{1}{x}\right)^{-1} \to +\infty,
\end{equation}
and the local shear stress is singular like
\begin{equation}\label{eq:tausing4_1}
    \tau\sim-\frac{2\mu_0\alpha}{3\beta \xcrit}\left(\frac{1}{\xcrit}-\frac{1}{x}\right)^{-1} \to -\infty
\end{equation}
as $x \to \xcrit^+$,
and hence the total force on the portion of the lower plate lying in the interval $\xcrit < x \le L$ is singular like
\begin{equation}\label{eq:Fsing4_1}
    F\sim-\frac{2\mu_0\alpha\xcrit}{3\beta}\log\left[-\frac{\alpha^2L\exp\left(-\frac{\beta p_L}{\mu_0}\right)}{6\beta U \xcrit \left(\frac{1}{\xcrit}-\frac{1}{x}\right)}\right] \to -\infty
\end{equation}
as $x \to \xcrit^+$.

On the other hand, in the case $U>0$, the pressure behaves like
\begin{equation}\label{eq:psing4_2}
    p\sim-\frac{\mu_0}{\beta}(-\log x) \to -\infty,
\end{equation}
the viscosity behaves like
\begin{equation}\label{eq:musing4_2}
    \mu\sim\frac{\mu_0\alpha^2 x}{6\beta U} \to 0^+,
\end{equation}
and the local shear stress behaves like
\begin{equation}\label{eq:tausing4_2}
    \tau\sim\frac{2\mu_0\alpha}{3\beta}
\end{equation}
as $x \to 0^+$,
and hence the total force is given by
\begin{equation}\label{eq:Fsing4_2}
    F=\frac{4\mu_0U}{\alpha}\left\{\exp\left(-\frac{\beta p_L}{\mu_0}\right)-\frac{6\beta U}{\alpha^2 L}\right\}^{-1}\log\left[\frac{\alpha^2L}{6\beta U}\exp\left(-\frac{\beta p_L}{\mu_0}\right)\right].
\end{equation}

% Section 3
\section{Conclusions}

The results given in Sections \ref{sec:sec2_2}--\ref{sec:sec2_4} show that, at least for the three simple models for the dependence of viscosity on pressure considered in the present work, the effect of a pressure-dependent viscosity is qualitatively different for the classical viscous scraper and the reverse viscous scraper problems.
Specifically, for the classical scraper (i.e., a drag in) problem the high pressure, and hence the high viscosity of the fluid, that occur near the point of contact between the plates (i.e., in the limit $x \to 0^+$) strengthen the non-integrable singularity in the shear stress that occurs for a fluid with constant viscosity, and so, as in the case of a fluid with constant viscosity, mean that, at least in theory, an infinite force is required to move the lower plate.
On the other hand, for the reverse scraper (i.e., a drag out) problem, the singularity in the shear stress is removed, and so a finite force is sufficient to move the lower plate. However, this is achieved by reducing the viscosity of the fluid to zero at the point of contact between the plates which, while theoretically interesting, is expected to be outwith the range of physical validity of the models for the dependence of viscosity on pressure.
It would be interesting to investigate if similar effects occur in other lubrication flows in which high pressures can occur, such as the viscous squeeze-film problem.

\bibliographystyle{spmpsci}

\end{document}